\documentstyle[preprint,prc,aps,psfig]{revtex}
\begin{document}
\draft
\tighten
\newcommand{\beq}{\begin{eqnarray}}
\newcommand{\eeq}{\end{eqnarray}}
\title{Color superconducting quark matter core in the third family of 
compact stars }
\author{Sarmistha Banik and Debades Bandyopadhyay}
\address{Saha Institute of Nuclear Physics, 1/AF Bidhannagar,
Kolkata 700 064, India}

\maketitle

\begin{abstract}
We investigate first order phase transitions from $\beta$-equilibrated 
hadronic matter to color flavor locked quark matter in compact star interior. 
The hadronic phase including hyperons and Bose-Einstein condensate of 
$K^-$ mesons is described by the relativistic field theoretical model with 
density dependent meson-baryon couplings. The early appearance of hyperons 
and/or Bose-Einstein condensate of $K^-$ mesons
delays the onset of phase transition to higher density. In the presence of 
hyperons and/or $K^-$ condensate, the overall equations of state become softer 
resulting in smaller maximum masses than the cases without hyperons and $K^-$ 
condensate. We find that the maximum mass
neutron stars may contain a mixed phase core of hyperons, $K^-$ condensate
and color superconducting quark matter. 
Depending on the parameter space, we also observe that there is a stable
branch of superdense stars called the third family branch beyond the neutron 
star branch. Compact stars in the third family branch may contain pure color 
superconducting core and have radii smaller than those of  the neutron star 
branch. Our results are compared with the recent observations on 
RX J185635-3754 and the recently measured mass-radius 
relationship by X-ray Multi Mirror-Newton Observatory.
\end{abstract}
\bigskip

\section{Introduction}
The theoretical investigation of mass-radius relationship of compact stars is
important because those could be directly compared with measured masses and
radii from various observations. Consequently, the composition and equation of
state (EoS) of dense
matter in neutron star interior might be probed. Recently, a lot of interest
has been generated about an isolated neutron star (INS) known as
RX J185635-3754 \cite{Pon}.
Several interesting features of this INS have emerged from various observations
by Chandra X-ray observatory and Hubble Space
Telescope (HST). It is one of the closest isolated neutron stars to our Sun. It
provides a great opportunity to measure its radius and put important
constraints on the EoS of dense matter in its core because of its relative
brightness, isolated nature and above all a thermal spectrum from X-ray to
optical wave lengths. Already, one group who has analysed the Chandra data
using a featureless blackbody spectrum, has claimed its radius is
$\sim 8$ km or less \cite{Dra}. This implies a very soft EoS including
exotic forms of matter. And their prediction is the compact star might be a
quark star \cite{Dra}. Alternatively, the Chandra data have been interpreted 
using a neutron star model where the gaseous outer layers of the neutron star
turn into a solid due to a phase transition and this model gives an apparent 
radius 
$\sim 10-12$ km \cite{Zan}. On the other hand, the analysis of HST data on that
compact star indicated a radius R=$11.4 \pm  2$ km and a mass M=$1.7 \pm 0.4
M_{solar}$ which may be explained by many EoS including hyperons 
and antikaon condensate \cite{Lat02}. Recently three strong spectral lines 
in the spectra of 28 bursts of low mass X-ray binary EXO0748-676 have been
observed by X-ray Multi Mirror-Newton Observatory. The estimated gravitational 
red shift from those lines is $z=0.35$ \cite{Cot}. This provides 
us with many important informations about the structure of compact star 
because the gravitational redshift depends on the mass-radius ratio of a compact
star.

The composition and structure of compact stars depend on the nature of strong 
interaction.
Neutron star matter encompasses a wide range of densities- the density of iron
nucleus at the surface to several times normal nuclear matter density in the 
core. Different phases of exotic matter with large strangeness fraction such as
hyperon matter \cite{Gle97}, Bose-Einstein condensates of strange mesons 
\cite{Kap,Pra97,Sch96,Gle99,Pal,Bani00,Bani01,Bani02} and 
quark matter \cite{Gle97,Pra97,Far84,Schr} may
occur in neutron star interior. It was extensively investigated how the 
appearance of various forms of exotic matter influences the EoS 
and mass-radius relationship of compact stars. It is now well understood that
each exotic component of dense matter makes the EoS soft. It was found in 
theoretical
investigations \cite{Lat01} that a soft EoS generally gave rise to a compact 
star with smaller maximum mass and radius than those of a stiffer EoS.

A transition from hadronic matter to deconfined strange quark matter is a 
possibility in neutron star interior. In earlier investigations, the quark 
phase was described
by the MIT bag model \cite{Gle97,Far84} in the hadron-quark phase transition. 
The recent development in 
dense matter physics points to the fact that the quark matter may be a 
color superconductor \cite{Bar77,Frau,Bai,Alf98,Rap,Alf99,Ris}. In this case, 
quarks near their Fermi surfaces form Cooper
pairs because quark-quark interaction is attractive in the antisymmetric 
color channel. The formation of diquark condensates breaks the color gauge 
symmetry. It was shown that at very high density quarks with all three flavors 
and  all three colors might pair up so as to produce an energetically 
favored state called the color-flavor-locked (CFL) phase\cite{Raj01}. The 
color neutrality constraint is to be imposed in the CFL quark matter 
because a macroscopic chunk of quark matter must
be color singlet \cite{Raj02,Stei02}. Consequently, the CFL quark matter is 
charge neutral. Color and charge neutrality in the CFL 
quark matter require a non-zero value for the color
chemical potential and electron chemical potential $\mu_e=0$ 
respectively \cite{Raj02,Stei02}. As the CFL condensate breaks chiral 
symmetry, the lightest degrees of freedom
in this phase are Nambu-Goldstone bosons \cite{Raj01,Red01,Bed02}. 
It was shown by various groups how the symmetric CFL phase behaves under 
stresses such as non-zero 
strange quark mass and electron chemical potential \cite{Red01,Bed02}. The CFL 
phase could relax under those stresses forming meson condensation. One such 
possibility is $K^0$ condensation in the charge neutral CFL quark matter 
\cite{Red01}. 

Recently, nuclear-CFL quark matter phase transition 
\cite{Alf01} and its impact on the structure of compact stars have been 
studied \cite{Alf02}. Also, the structure of compact stars including pure CFL 
quark matter has been studied by others \cite{Hov02}. Along with CFL quark 
matter, hyperons and $K^-$ condensate could exist in compact star interior.
Many interesting things may happen if one form of exotic matter sets in before 
the other. The early appearance of hyperons was found to delay the 
hadron-unpaired quark matter phase transition or vice-versa \cite{Pra97}. 
Similar effects of 
hyperons on the threshold of $K^-$ condensation in hadronic matter was observed 
by various groups \cite{Pra97,Pal,Bani00,Bani01,Bani02}. Now the question is
what could happen to nuclear-CFL quark matter phase transition if hyperons and
$K^-$ condensate appear in compact star matter. So far no calculation of 
compact stars involving color superconducting quark matter, hyperons and $K^-$
condensate has been performed. In this paper, we 
investigate how the formation of hyperons and antikaon condensate in the 
hadronic matter influences the phase transition from hadronic matter to the CFL 
quark matter including $K^0$ condensate and the structure of compact stars. 
The paper is organised in the following way. In Sec. II, we describe the DDRH
model for hadronic phase and also the CFL quark matter. Results of our 
calculation are explained in Sec. III. And Sec. IV provides a summary and 
conclusions.

\section{Formalism}
Here, we discuss a phase transition from hadronic matter to the CFL quark 
matter in compact stars. The $\beta$-equilibrated and charge neutral bulk 
hadronic phase is described within the framework of a
Density Dependent Relativistic Hadron (DDRH) model 
\cite{Bani02,Hof1,Hof2,Fuc95,Len95}. In 
the DDRH model, many body correlations are taken into account by density 
dependent meson-baryon couplings. The hadronic phase is composed of all species
of the baryon octet, (anti)kaons, electrons and muons. 
Therefore, the total Lagrangian density in the hadronic phase is written as 
${\cal L} = {\cal L}_B + {\cal L}_K + {\cal L}_l$. In the DDRH model, 
baryon-baryon interaction is given by the  Lagrangian density (${\cal L}_B$)
\cite{Hof2},
\begin{eqnarray}
{\cal L}_B &=& \sum_B \bar\Psi_{B}\left(i\gamma_\mu{\partial^\mu} - m_B
+ g_{\sigma B} \sigma - g_{\omega B} \gamma_\mu \omega^\mu 
- \frac{1}{2} g_{\rho B} 
\gamma_\mu{\mbox{\boldmath $\tau$}}_B \cdot 
{\mbox{\boldmath $\rho$}}^\mu +\frac {1} {2} g_{\delta B}{\mbox{\boldmath 
$\tau$}}_B \cdot {\mbox{\boldmath $\delta$}} \right)\Psi_B\nonumber\\
&& + \frac{1}{2}\left( \partial_\mu \sigma\partial^\mu \sigma
- m_\sigma^2 \sigma^2\right) + \frac{1}{2}\left( \partial_\mu \delta\partial
^\mu \delta
- m_\delta^2 \delta^2\right)  
 -\frac{1}{4} \omega_{\mu\nu}\omega^{\mu\nu}\nonumber\\
&&+\frac{1}{2}m_\omega^2 \omega_\mu \omega^\mu
- \frac{1}{4}{\mbox {\boldmath $\rho$}}_{\mu\nu} \cdot
{\mbox {\boldmath $\rho$}}^{\mu\nu}
+ \frac{1}{2}m_\rho^2 {\mbox {\boldmath $\rho$}}_\mu \cdot
{\mbox {\boldmath $\rho$}}^\mu.
\end{eqnarray}

The Lagrangian density for (anti)kaon condensation in the minimal coupling 
scheme \cite{Gle99,Bani02} is
\begin{equation}
{\cal L}_K = D^*_\mu{\bar K} D^\mu K - m_K^{* 2} {\bar K} K ~,
\end{equation}
where the covariant derivative
$D_\mu = \partial_\mu + ig_{\omega K}{\omega_\mu} + ig_{\rho K}
{\mbox{\boldmath $\tau$}}_K \cdot {\mbox{\boldmath $\rho$}}_\mu/2$
and the effective mass of (anti)kaons 
$m_K^* = m_K - g_{\sigma K} \sigma + \frac {1} {2} g_{\delta K} \delta$.
Unlike meson-baryon couplings, meson-(anti)kaon couplings are 
density-independent here. The threshold condition of $K^-$ condensation is
$\omega_{K^-} = \mu_e $, where $\omega_{K^-}$ is the in-medium energy of $K^-$
mesons for s-wave condensation \cite{Bani02}.

The Lagrangian density for leptons is
\begin{eqnarray}
{\cal L}_l &=& 
\sum_l \bar\psi_l\left(i\gamma_\mu {\partial^\mu} - m_l \right)\psi_l ~,
\end{eqnarray}
where $\psi_l$ ($l \equiv {e, \mu}$) is lepton spinor. 

The charge neutrality condition in the bulk hadronic phase is 
\begin{equation}
Q^h = \sum_b q_b n^h_b -n_{K^-}-n_e -n_\mu =0,
\end{equation}
where $q_b$ and $n_b^h$ are electric charge and the number density of baryon b 
in the pure hadronic phase, respectively and $n_{K^-}$,
$n_e$ and $n_\mu$ are number densities of $K^-$, electrons and muons 
respectively. The energy density ($\varepsilon^h$) and the pressure
($P^h$) in the hadronic phase are given by Ref.\cite{Bani02}.

The pure CFL quark matter is composed of paired quarks of all flavors and 
colors and neutral kaons which are Goldstone bosons arising due to the 
breaking of chiral symmetry in the CFL phase. The thermodynamic potential
for electric and color charge neutral CFL quark matter to order $\Delta^2$ 
is given by \cite{Raj01,Alf01,Alf02}
\begin{eqnarray}\label{ome}
\Omega^q_{CFL}={6 \over \pi^2} \int_0^\nu p^2 (p-\mu) dp 
+{3 \over {\pi^2}}\int_0^\nu
p^2 ( \sqrt{p^2 +m_s^2} - \mu) dp -{3 \Delta^2 \mu^2 \over \pi^2}+B~,
\end{eqnarray}
where $\Delta$ is the color superconducting gap and B is the bag contribution. 
The first two terms of 
Eq. (\ref{ome}) give the thermodynamic potential of (fictional) unpaired quark 
matter in which all quarks that are going to pair have a common Fermi momentum 
$\nu$ which minimizes the thermodynamic potential of the fictional unpaired 
quark matter \cite{Raj01,Alf01,Alf02}.  
The third term is the contribution of the CFL condensate to $\Omega^q_{CFL}$. 
The common Fermi momentum is 
\begin{eqnarray}\label{fer}
\nu = 2 \mu - \sqrt {\mu^2 + {m_s^2 \over 3}}~,
\end{eqnarray}
where $\mu$ is the average quark chemical potential and $m_s$ is the strange 
quark mass. Studying the pairing ansatz in the CFL phase, it was shown by 
various authors \cite{Stei02} that 
\beq
n_u=n_r, \qquad n_d=n_g  \qquad\mbox{and}\qquad n_s=n_b~,
\eeq
where $n_r$, $n_g$, $n_b$ 
and $n_u$, $n_d$, $n_s$ are color and flavor number densities respectively. 
It follows from the above relation that color neutrality automatically enforces 
electric charge  neutrality in the CFL phase. The quark number densities
are $n_u=n_d=n_s= {(\nu^3+2 \Delta^2 \mu)}/{\pi^2}$. As the color 
neutral
CFL quark matter is electric charge neutral, the corresponding electric 
charge chemical potential is $\mu_e=0$. Similarly, the color chemical potential
 corresponding to color charge $T_3$ is $\mu_3=0$. However, the color chemical
potential for color charge $T_8$ is $\mu_8 \neq 0$ to maintain color charge 
neutrality in the CFL phase \cite{Raj02,Stei02}.

The thermodynamic potential ($\Omega^{K^0}_{CFL}$) due to $K^0$ condensate is 
given by Ref.\cite{Red01}. The pressure in the CFL phase  including
$K^0$ condensate is given by $P^q=-\Omega_{CFL}^q-\Omega_{CFL}^{K^0}$. The
energy density ($\epsilon^q$) in the CFL phase is obtained from the 
Gibbs-Duhem relation \cite{Mad}. 

We now describe the mixed phase of the above mentioned two phases because the 
phase transition from hadronic to the 
CFL quark matter is a first order phase transition. Earlier 
the mixed phase of hadronic and unpaired quark matter was studied by various 
authors \cite{Gle97,Schr,Gle92}. In that case, baryon number and electric 
charge were two conserved charges in individual bulk phase. Glendenning argued 
conserved charges may be shared by two phases in equilibrium in the 
coexistence phase and the mixed  phase is to be determined by Gibbs conditions 
along with global baryon and electric charge conservation laws \cite{Gle92}.
It was found that the net positive charge of the hadronic phase was neutralised 
by the net negative charge of the unpaired quark matter in the mixed phase. 
Unlike unpaired quark matter, electric charge is not present in the CFL quark 
matter. And baryon and color charge are conserved quantities in the CFL
phase. Recently, the mixed phase of nuclear and CFL quark matter including 
Goldstone bosons $\pi^-$ or $K^-$ was constructed using Glendenning's 
prescription \cite{Alf01,Alf02}. It was shown there that $K^-$ or 
$\pi^-$ condensate was formed at the cost of electrons of the hadronic phase 
and made the CFL phase negatively charged in the mixed phase. 
As color charge in the bulk hadronic phase and
electric charge in the bulk CFL phase can not be present, we relax
the global conservation laws of Glendenning \cite{Gle92} and
consider only local electric charge neutrality
in the hadronic phase and local color charge neutrality in the CFL phase. Here
the mixed phase is determined by Gibbs phase rules and the global conservation 
law for baryon number. The Gibbs conditions read
\begin{eqnarray}
P^h&=& P^q~,\\
\mu_n& =& 3 \mu~,
\end{eqnarray}
where $\mu_n$ and $\mu$ are neutron and quark chemical potential. The global
baryon number conservation law is imposed through the relation
\begin{equation}
n_b=(1-\chi) n_b^h + \chi n_b^q~,
\end{equation}
where $\chi$ is the volume fraction of the CFL phase in the mixed
phase and $ n_b^h $ and $n_b^q$ are baryon densities in the hadronic and CFL
phase respectively. The total energy density in the mixed phase is 
\begin{equation}
\epsilon=(1-\chi)\epsilon^h + \chi \epsilon^q~.
\end{equation}

\section{Results and Discussions}
In the DDRH model, the dependence of meson-nucleon vertices on total baryon
density is obtained from microscopic Dirac-Brueckner calculations of symmetric 
and asymmetric matter using Groningen nucleon-nucleon potential 
\cite{Len98,Len2,Mal}. We adopt a
suitable parameterisation for density dependent couplings as was given by Ref.
\cite{Hof1}.
The density dependent meson-nucleon couplings at saturation density are 
listed 
in Table I of Ref. \cite{Bani02}. The saturation properties calculated using 
these couplings are 
binding energy E/A=-15.6 MeV, saturation density $n_0 = 0.18 {\em fm^{-3}}$, 
symmetry energy coefficient $a_{sym} = 26.1$ MeV and incompressibility K = 282 
MeV \cite{Bani02}.

The density dependence of meson-hyperon vertices are obtained from density 
dependent meson-nucleon couplings using hypernuclear data \cite{Sch96} 
and scaling law \cite{Kei}. The 
ratio of meson-hyperon coupling to meson-nucleon coupling for Groningen 
potential is given by Table \ref{tab2} of Ref. \cite{Bani02}. Similarly, 
we employ the density 
independent meson-(anti)kaon couplings of Ref.\cite{Bani02} in our calculation.

Now we report results of our calculation. We have performed calculations for 
compact star matter allowing first order phase transition from i) nuclear to 
CFL quark matter (np$\rightarrow$ CFL+$K^0$), ii) nuclear+$K^-$condensate 
to CFL quark matter (np$K^-$$\rightarrow$ CFL+$K^0$), iii) hyperonic matter 
to CFL quark matter (np$\Lambda\Xi$$\rightarrow$ CFL+$K^0$), and iv) hadronic 
matter including hyperons and $K^-$ condensate to CFL quark matter
(np$\Lambda\Xi$$K^-$$\rightarrow$ CFL+$K^0$). In this calculation, we have
neglected the density dependence of strange quark mass ($m_s$), color 
superconducting gap ($\Delta$) and bag constant ($B^{1/4}$). 

First we discuss np$\rightarrow$ CFL+$K^0$ (denoted hereafter by NQ) matter
phase transition for different values of gap 
and strange quark mass whereas the bag constant is fixed at 
$B^{1/4} = 180$ MeV. We also show the results of nuclear to unpaired 
($\Delta =0$) quark matter phase transition. 
Nuclear matter is composed of neutrons, protons, electrons and muons 
whereas the CFL phase contains paired quarks and $K^0$ condensate.
For nuclear to unpaired quark matter phase transition, 
we have imposed both global baryon and electric charge conservation in the 
mixed phase. On the other hand, we have global baryon number conservation and 
local color and electric charge neutrality in the mixed phase of nuclear-CFL 
phase transition. The lower and upper boundary of the mixed phase for 
$\Delta =$ 0, 57 and 100 MeV  and $m_s = $ 150 and 200 MeV and 
$B^{1/4} = 180$ MeV are recorded in Table \ref{tab1}. The onset of phase
transition is at $u_l = n_l/n_0$ and the pure CFL phase begins at 
$u_u = n_u/n_0$. Here we note that the phase transition to quark matter 
is delayed to higher densities for a larger value of $m_s$. On the other hand, 
the phase transition occurs earlier for case $\Delta = 100$ MeV than that of 
case $\Delta = $ 0 and 57 MeV irrespective of strange quark mass. 
We also note that the extent of the mixed phase involving unpaired
quark matter is the largest among all the cases studied here. The average 
quark chemical potential at the onset of the phase transition is shown in 
Table I. We find that the CFL phase is energetically favoured over unpaired
quark matter because $\Delta > {m_s^2} / {4\mu}$ is satisfied \cite{Raj02}.  

\begin{figure} 
\begin{center}
\psfig{figure=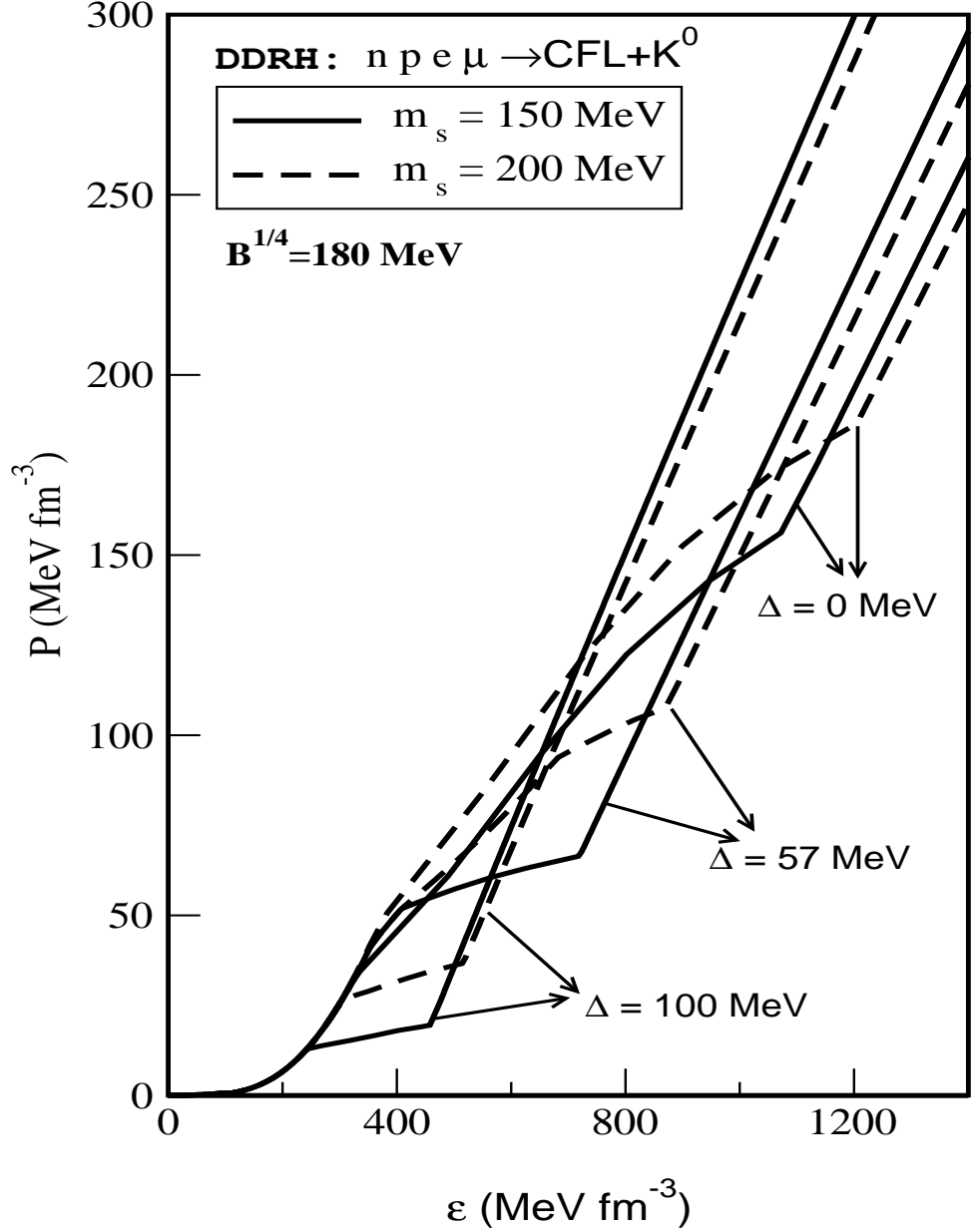,width=16cm,height=19cm}
\caption{The equations of state, pressure $P$ versus energy density 
$\varepsilon$ for  n, p,  
lepton and CFL quark matter including $K^0$ condensate
with $B^{1/4} = 180$ MeV and different values of $\Delta$=0 (unpaired quark matter), 57 MeV and 100 MeV and $m_s$= 150 MeV and 200 MeV are compared.
}\label{eos}
\end{center}
\end{figure}

\begin{figure} 
\begin{center}
\psfig{figure=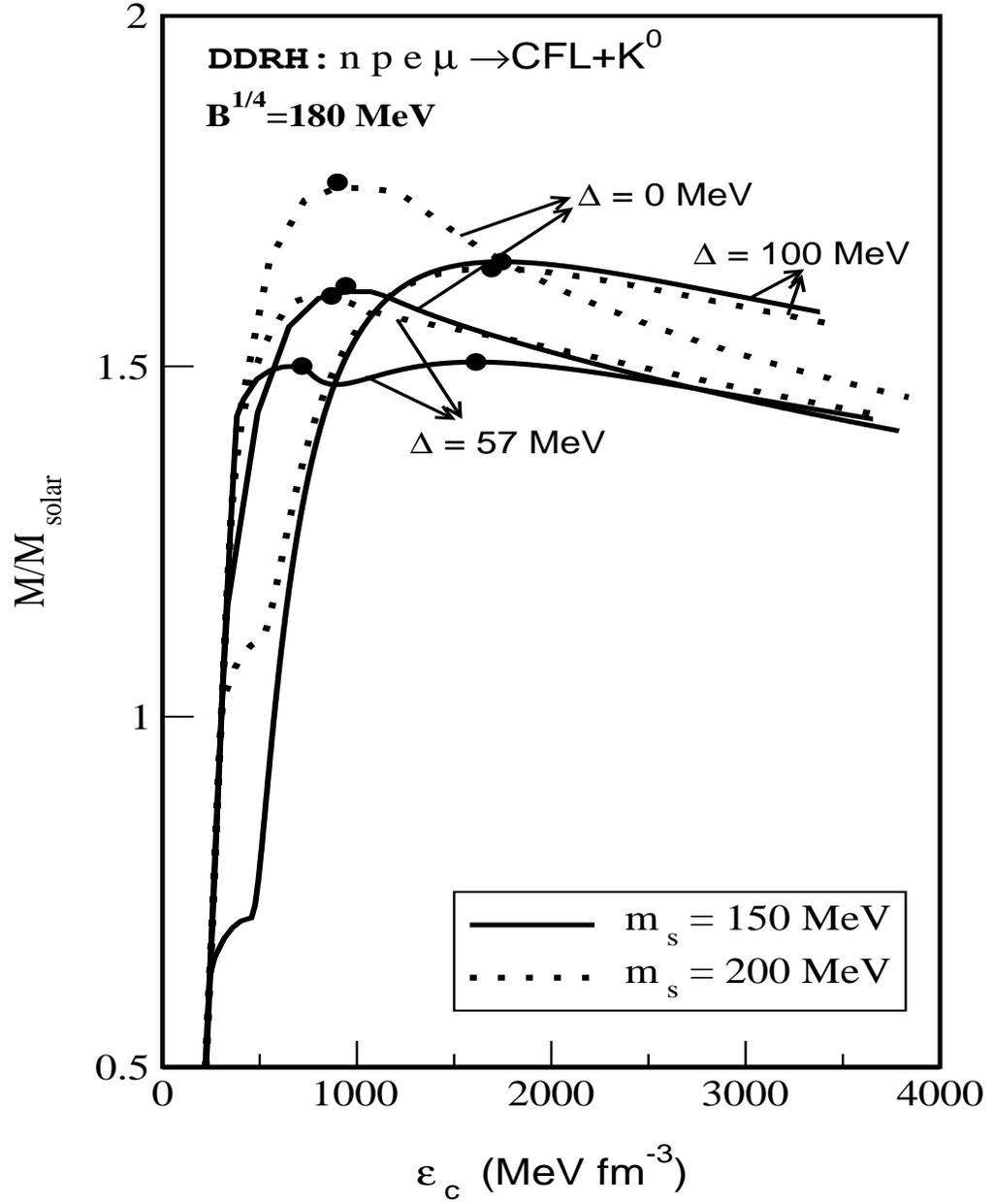,width=16cm,height=19cm}
\caption{The corresponding mass sequences for EoS of Fig. \ref{eos} are plotted . }\label{mase}
\end{center}
\end{figure}

Pressure versus energy density and gravitational mass against central
energy density for compact stars with NQ matter and different values of 
gap ($\Delta = 0, 57$ and $100$ MeV), $m_s$ = 150 and 200 MeV and 
$B^{1/4} = 180$ MeV are exhibited in Fig. \ref{eos} and Fig. \ref{mase} 
respectively. The static structures of spherically symmetric neutron 
stars calculated using Tolman-Oppenheimer-Volkoff (TOV) equations and the 
equations of state of Fig. \ref{eos}, are presented in Fig. \ref{mase}. 
We adopt Negele and Vautherin \cite{Neg} and Baym-Pethick-Sutherland 
\cite{Bay} equations of state to describe nuclear matter at very low density. 
The maximum neutron star masses ($M/M_{solar}$) and their corresponding central
densities ($u_{cent} = n_{cent}/n_0$) are listed in Table I. 
For unpaired quark matter and paired quark matter with $\Delta = 57$ MeV, the 
overall EoS with $m_s=200$ MeV is stiffer than that of 
$m_s = 150$ MeV. Consequently, the compact star in the former case has a
larger maximum mass than that of the latter case. The effect of $m_s$ is 
pronounced for unpaired quark matter. However,  
for $\Delta = 100$ MeV and different values of $m_s$ there is no significant 
change in the EoS and maximum star masses. 
Again, we find the maximum
masses for $\Delta = 57$ MeV are reduced compared with those for $\Delta = 100$ 
MeV because the EoS in the latter case is stiffer. From Table I, it is observed that compact stars have pure CFL quark matter core for $\Delta = 100$ MeV
whereas compact stars, in other cases, contain a mixed phase core of nuclear 
and unpaired or CFL quark matter. We note that $K^0$ condensate in the CFL
phase does not contribute significantly in the energy density and pressure. 
Recently, Alford and Reddy
\cite{Alf02} also investigated the influence of $m_s$ and $\Delta$ on the EoS 
and structure of compact stars in nuclear-CFL quark matter phase transition 
using different models for the hadronic phase. Our results for stars including
NQ matter are in qualitative agreement with their findings.
In our calculation, we do
not consider the variation of bag constant on nuclear-CFL quark matter phase
transition. But we expect similar qualitative feature as it was observed
for the phase transition involving unpaired quark matter. It is worth 
mentioning here that a larger value of bag constant makes the EoS of unpaired
quark matter soft delaying the phase transition to higher density \cite{Pra97}.
 
In earlier investigations, it was shown that exotic components of matter such as
hyperons \cite{Gle97}, Bose-Einstein condensate of $K^-$ mesons 
\cite{Pal,Bani00,Bani01,Bani02} and quarks \cite{Pra97} could appear in
$\beta$-equilibrated and electric charge neutral matter around (2-4)$n_0$. It
was also studied how hyperons delayed the onset of antikaon condensation 
\cite{Pal,Bani00,Bani01,Bani02} and the phase transition to unpaired quark
matter \cite{Pra97} to higher density. With the appearance of each exotic 
phase in dense matter, the EoS becomes softer. Here we perform
calculation involving hyperons, antikaon condensate and quarks and examine 
how exotic phases of matter compete with each other in compact star interior.
Also, we explore whether there is any window in the parameter space for which 
all three exotic components of matter may coexist in the core of compact stars 
and its implication on their structures. 
   
In Table II, we show the mixed phase boundaries for compact star matter that
undergoes first order phase transition from 
np$\Lambda\Xi$$\rightarrow$ CFL+$K^0$ (denoted by NHQ), 
np$K^-$$\rightarrow$ CFL+$K^0$ (N$\bar K$Q) and 
np$\Lambda\Xi$$K^-$$\rightarrow$ CFL+$K^0$ (NH$\bar K$Q). These results are
obtained for strange quark mass $m_s = 150$ MeV. The early appearance of 
hyperons in NHQ star matter at 1.99$n_0$ gives rise to a softer EoS delaying 
the phase transition to CFL quark matter to higher densities for all values of 
bag constants and gap as are shown in Table II. On the other hand, for 
N$\bar K$Q matter with $B^{1/4} = 180$ MeV and antikaon optical potential 
depth $U_{\bar K} = -180$ MeV, the onset of $K^-$ condensation produces a 
softer EoS postponing the CFL phase with $\Delta = 30$ MeV to higher density
and the reverse happens for the CFL phase with $\Delta = 57$ MeV. In the
calculation of NH$\bar K$Q matter using $B^{1/4} = 180 (185)$ MeV, 
$\Delta = 30 (57)$ MeV and $U_{\bar K} = -180 (-160)$ MeV, hyperons and $K^-$
condensate appear before the phase transition to the CFL phase. In the other 
case with $B^{1/4} = 180$ MeV, $\Delta = 57$ MeV and $U_{\bar K} = -180$ MeV, 
$K^-$ condensate appear in the mixed phase. We also perform 
calculation for NHQ, N$\bar K$Q and NH$\bar K$Q matter using parameters 
$B^{1/4} = 180$ MeV, $\Delta = 100$ MeV $m_s = 150$ MeV and 
$U_{\bar K} = -180$ MeV. For all cases, we note that the early onset of CFL 
phase at 
1.43$n_0$ forbids hyperons and/or $K^-$ condensate to appear in compact star
matter. In the following two paragraphs, we discuss the compositions of compact
star matter and how exotic phases of dense matter compete with each other.

\begin{figure} 
\psfig{figure=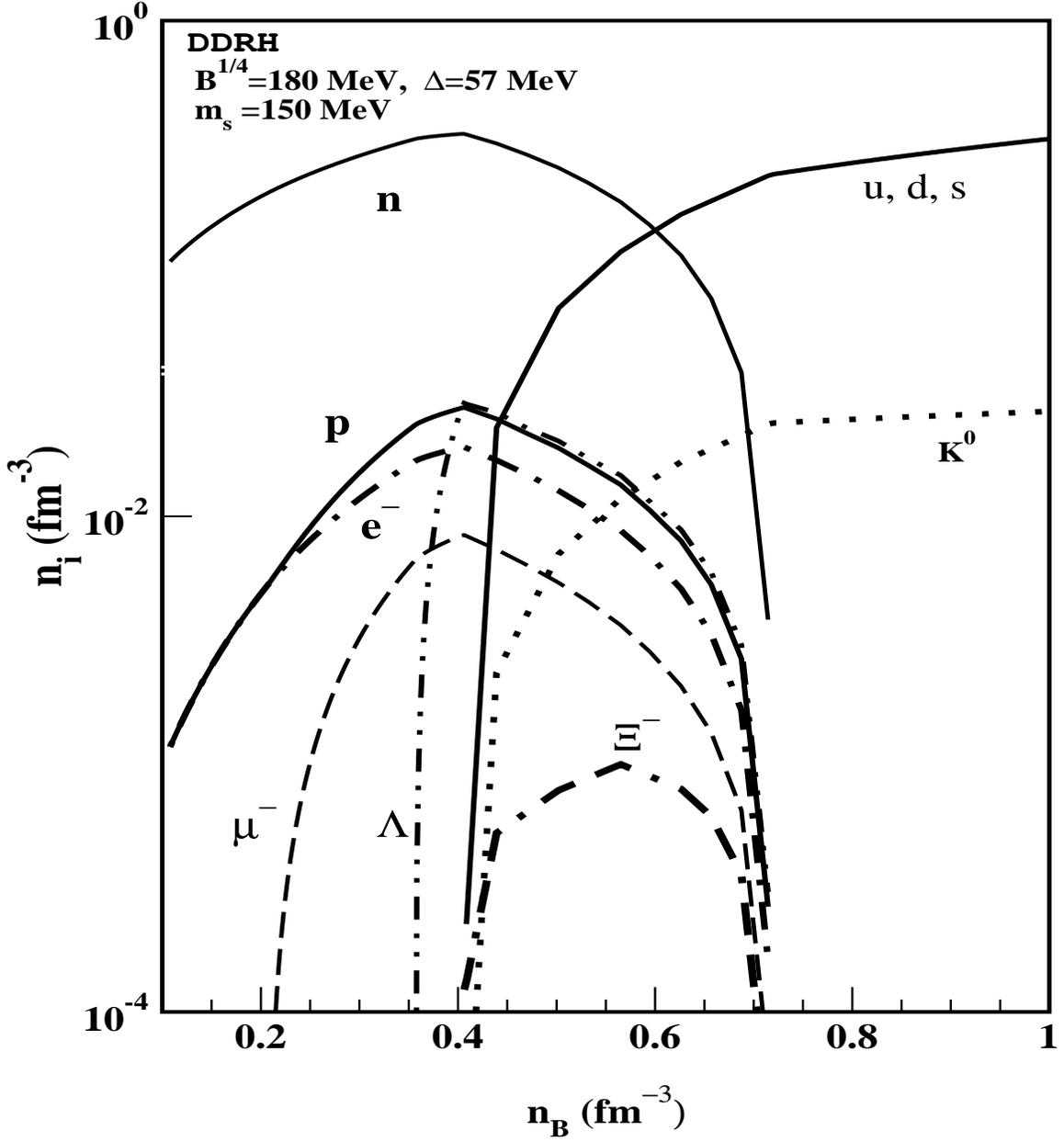,width=16cm,height=19cm}
\caption{Number densities ($n_i$) of various particles in
$\beta$-equilibrated n, p, $\Lambda$, $\Xi$, lepton and CFL+$K^0$
matter for $B^{1/4}=180$ MeV, $\Delta$= 57 MeV, 
and $m_s$= 150 MeV
as a function of baryon density.
}\label{hfrac}
\end{figure}

The composition of $\beta$-equilibrated NHQ matter relevant to compact stars
is displayed in Fig. \ref{hfrac}. The hadronic phase is composed of
n, p, $\Lambda$, $\Xi$ and leptons. On the other hand, we have
included the contribution of Goldstone boson $K^0$ in the CFL
phase. Here, we take bag constant $B^{1/4} = 180$ MeV, $m_s=150$
MeV and $\Delta$ = $57$ MeV to describe the CFL quark
matter. The charge neutrality in the hadronic phase is maintained by
protons, electrons and muons. We observe that $\Lambda$ hyperons
appear at $1.99n_0$, where $n_0=0.18$ ${\em fm^{-3}}$. The phase
transition from hadronic to CFL quark matter sets in at $2.27n_0$. 
The mixed phase is over at $3.97n_0$. It is noted here that negatively
charged $\Xi$ hyperons appear just before the onset of mixed phase at
$2.26n_0$. With the appearance of $\Xi^-$ hyperons, the density of electrons
drops fast. Also, neutral $K^0$ condensation appears with the onset of
CFL phase and it rises rapidly. 

\begin{figure} 
\psfig{figure=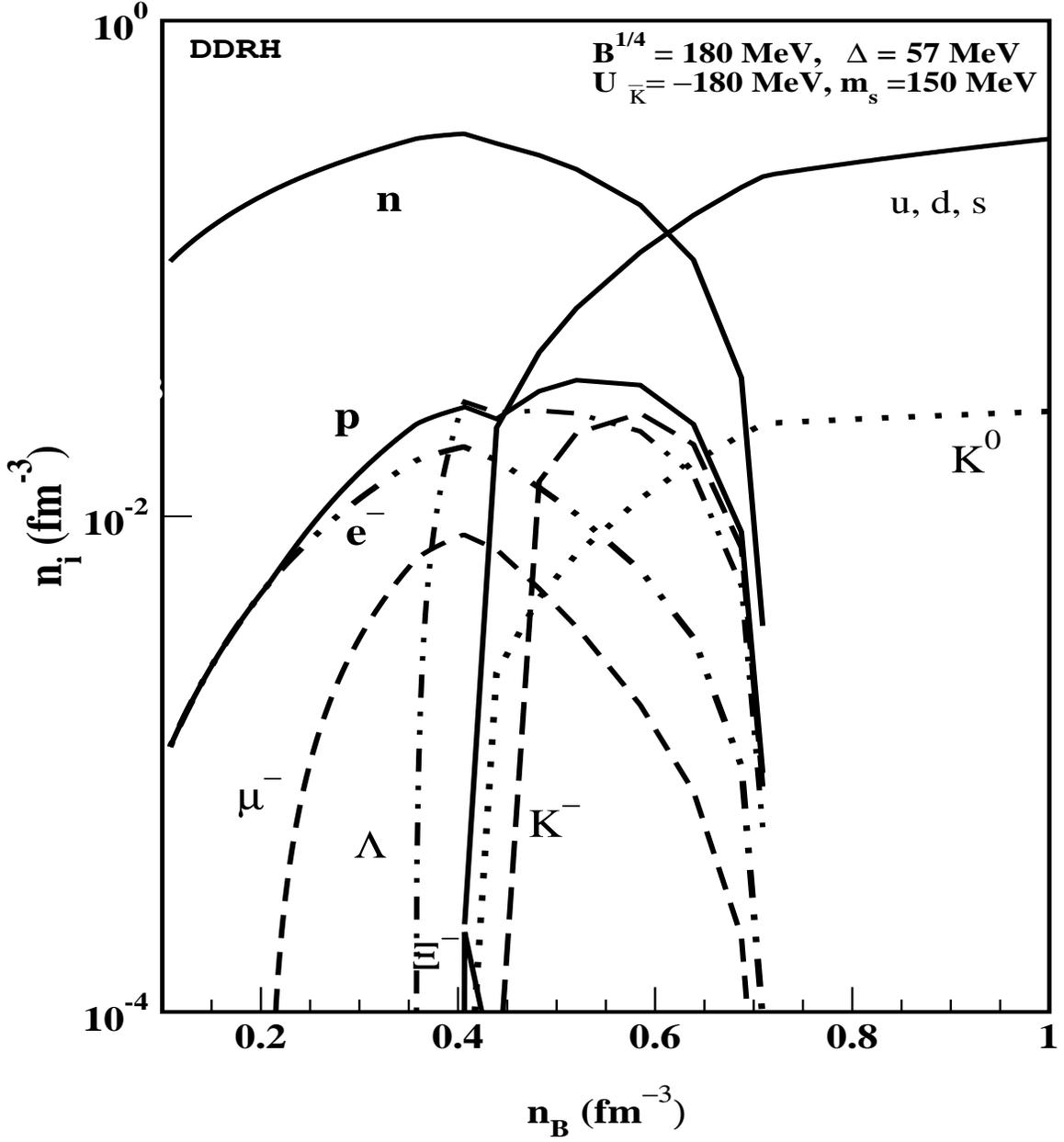,width=16cm,height=19cm}
\caption{Number densities ($n_i$) of various particles in
$\beta$-equilibrated n, p, $\Lambda$, $\Xi$, $K^-$, lepton and CFL+$K^0$
matter for $B^{1/4}=180$ MeV, $\Delta$= 57 MeV,  
$m_s$= 150 MeV
and  antikaon optical 
potential depth $U_{\bar K}(n_0) = -180$ MeV at normal nuclear matter 
density as a function of baryon density .
}\label{kfrac}
\end{figure}

Along with hyperons, we also investigate the role of $K^-$ condensation in the 
hadronic phase on the composition of NH$\bar K$Q star matter. In a recent 
calculation using DDRH model, it has been shown that $K^-$ condensation in 
hadronic matter is a second order phase transition \cite{Bani02}. In 
Fig. \ref{kfrac} we exhibit the particle abundances 
in $\beta$-equilibrated NH$\bar K$Q matter. 
The antikaon optical potential used for this calculation is 
$U_{\bar K} \left(n_0\right) = - 180$ MeV and bag constant, strange quark mass 
and gap are 180, 150 and 57 MeV respectively.
Here we find that the mixed phase begins at the same density point i.e.
$2.27n_0$ as it is noted in Fig. \ref{hfrac}. 
Before the onset of $K^-$ condensate, 
the charge neutrality is maintained by protons and leptons.
The condensation of $K^-$ mesons occurs in the mixed phase at $2.68n_0$. As soon
as $K^-$ condensate appears, it rapidly replaces electrons and muons and grows
fast. Consequently, proton density increases. Also, we observe that negatively
charged $\Xi$ hyperon disappears almost as soon as it appears. There is no 
influence 
of $K^-$ condensate on the extent of the mixed phase because $K^-$ condensate
appears in the mixed phase. 
 
\begin{figure} 
\begin{center}
\psfig{figure=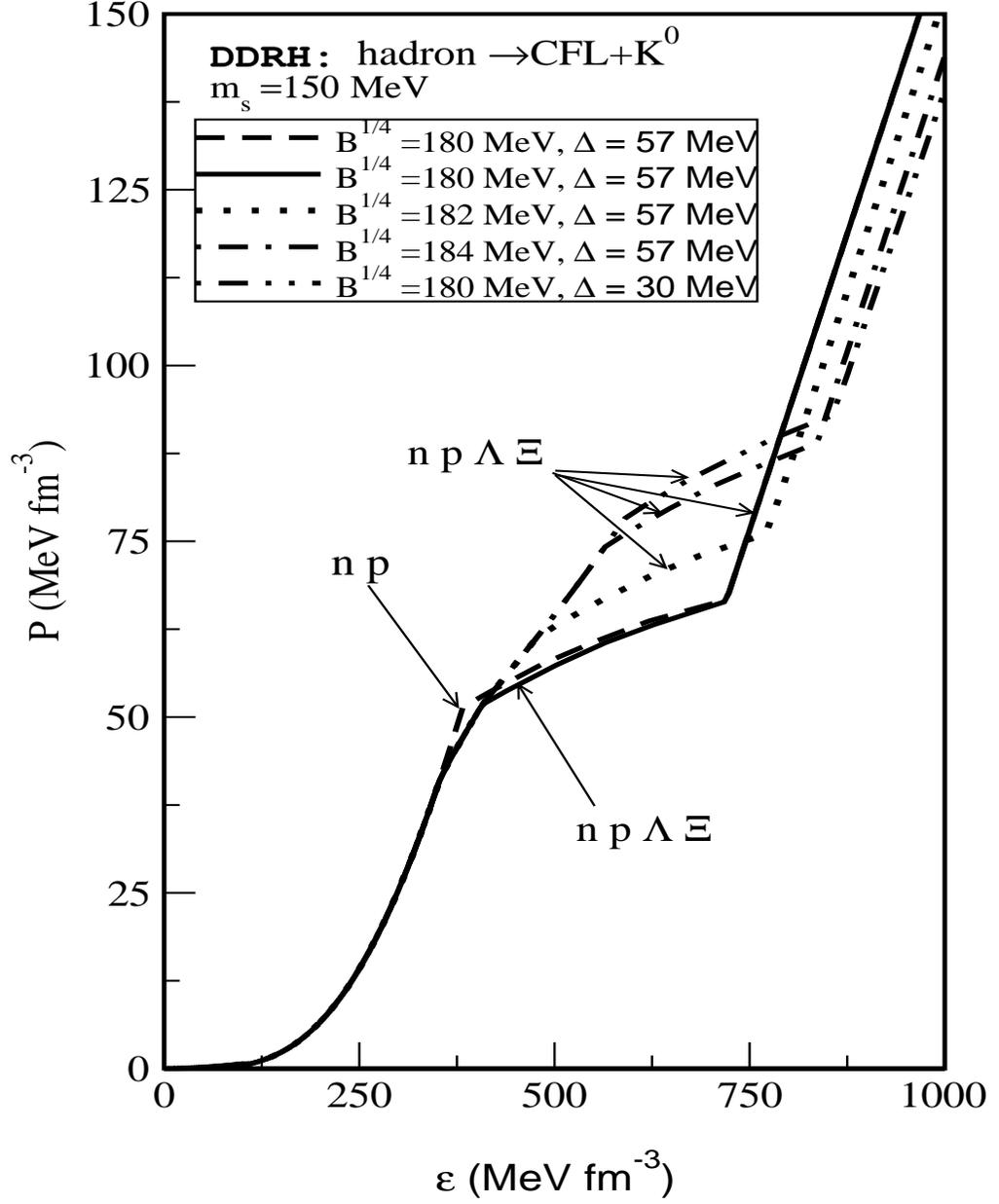,width=16cm,height=19cm}
\caption{The equation of state, pressure $P$ versus energy density 
$\varepsilon$
is shown here. The results are for  n, p,  
lepton and CFL+$K^0$  matter
(dashed line) and n, p, $\Lambda$, $\Xi$, lepton and CFL+$K^0$ 
matter (other lines)  for  
$m_s$= 150 MeV and different values of bag constant and gap.
}\label{eost}
\end{center}
\end{figure}

Equations of state (pressure versus energy density) for $\beta$-equilibrated 
and charge neutral NQ and NHQ compact star matter are plotted in 
Fig. \ref{eost}. 
The curves correspond to calculations for $m_s$ = 150 MeV and 
different values of bag constant and gap. 
The dashed line stands for the EoS of NQ matter whereas 
the other lines
represent those of NHQ matter for different parameter sets. For 
 $B^{1/4} = 180$ MeV and $\Delta = 57$ MeV, the EoS for NHQ 
matter is slightly
softer compared with the corresponding EoS for NQ matter. As the
bag constant is increased keeping the gap fixed, we find the corresponding 
equations of state become stiffer. The stiffest EoS among all cases considered 
here corresponds to the EoS for NHQ matter with $\Delta = 30$ MeV.

\begin{figure} 
\begin{center}
\psfig{figure=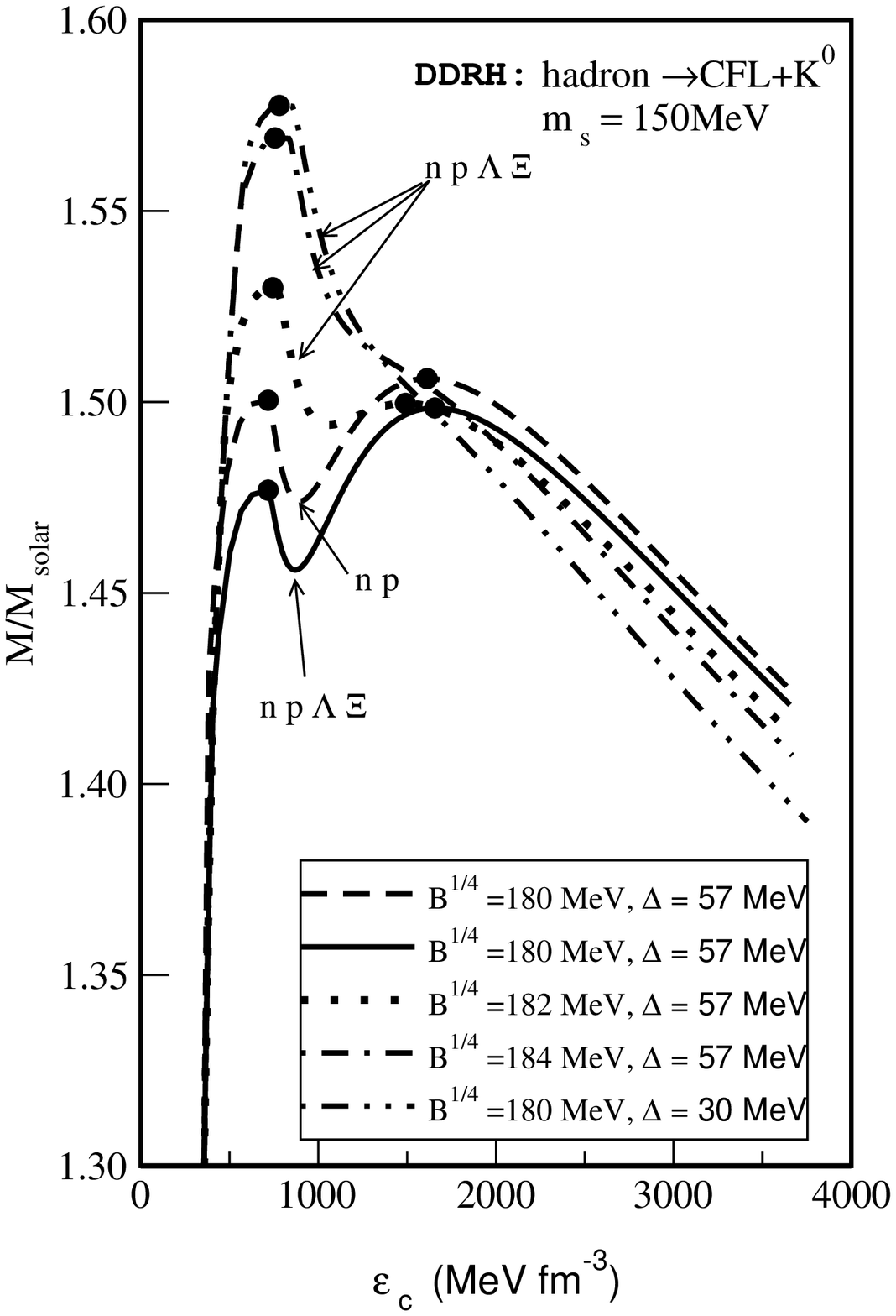,width=16cm,height=19cm}
\caption{The compact star mass sequences are plotted with central energy
density for the corresponding EoS of Fig. \ref{eost}.}\label{mast}
\end{center}
\end{figure}

The maximum neutron 
star masses ($M_{max}/M_{solar}$) and  their central densities ($u_{cent} = 
n_{cent}/n_0$) calculated with the above mentioned equations of state including
NQ and NHQ matter are listed in Table \ref{tab1} and Table \ref{tab2}. 
The static neutron star sequences representing the 
stellar masses 
$M/M_{solar}$ and the corresponding central energy densities 
($\varepsilon_c$) for different values of bag constant and gap
are shown in Fig. \ref{mast}. The closed circle on each curve denotes the 
maximum mass star. 
For $B^{1/4} = 180$ MeV and $\Delta=57$ MeV, the maximum mass
of NQ star is $1.500M_{solar}$ corresponding to central energy 
density
717.46 ${\em MeV/fm^3}$, whereas that of NHQ star is $1.477M_{solar}$ 
corresponding to $\varepsilon_c=$ 717.47 ${\em MeV/fm^3}$. The maximum mass of 
the  star including hyperons is smaller than that of the star with NQ matter 
because the EoS is softer in former case. The maximum masses of compact 
stars increase as we increase the bag constant or decrease the gap. This
is evident from Fig. \ref{mast} and Table \ref{tab2}. We also note from 
Table II that the central densities corresponding to maximum masses of 
neutron stars including NHQ matter fall in the mixed phase. 
Consequently the maximum mass neutron stars contain a hyperon-CFL 
quark mixed phase core. For each of the  bottom three curves in
Fig. \ref{mast}, we find an unstable region followed by a 
stable sequence of superdense stars beyond the neutron star branch. The stable
branch of superdense stars beyond the neutron star branch is called the third 
family branch \cite{Bani01,Schr,Uli,Glen00,Sch02,Fraga}. It was earlier shown 
by various authors that the compact stars in
the third family branch had different compositions and smaller radii than those
of the neutron star branch. In this calculation, the superdense stars in the 
third family branch may contain a pure CFL quark matter core because those 
stars appear after the mixed phase is over.

\begin{figure} 
\begin{center}
\psfig{figure=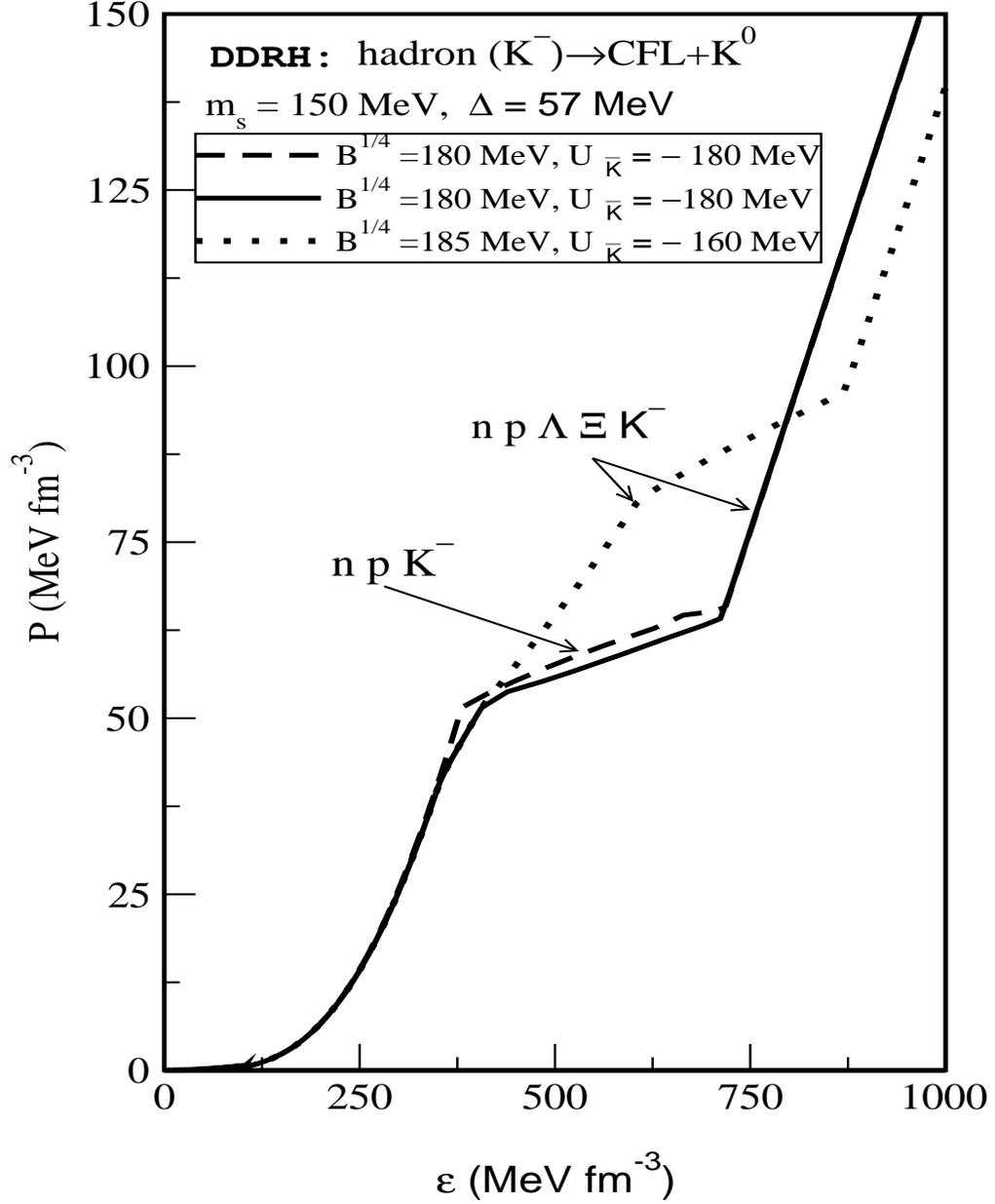,width=16cm,height=19cm}
\caption{The equation of state, pressure $P$ versus energy density 
$\varepsilon$,
is shown here. The results are for hadronic matter including 
$K^-$ condensate
and CFL+$K^0$  matter
for  
$m_s$= 150 MeV, $\Delta$=57 MeV and different values of bag constant and 
antikaon optical potential depth $U_{\bar K}(n_0)=-160$ and $-180$ MeV at 
normal nuclear matter density. }\label{eosk}
\end{center}
\end{figure}

\begin{figure} 
\begin{center}
\psfig{figure=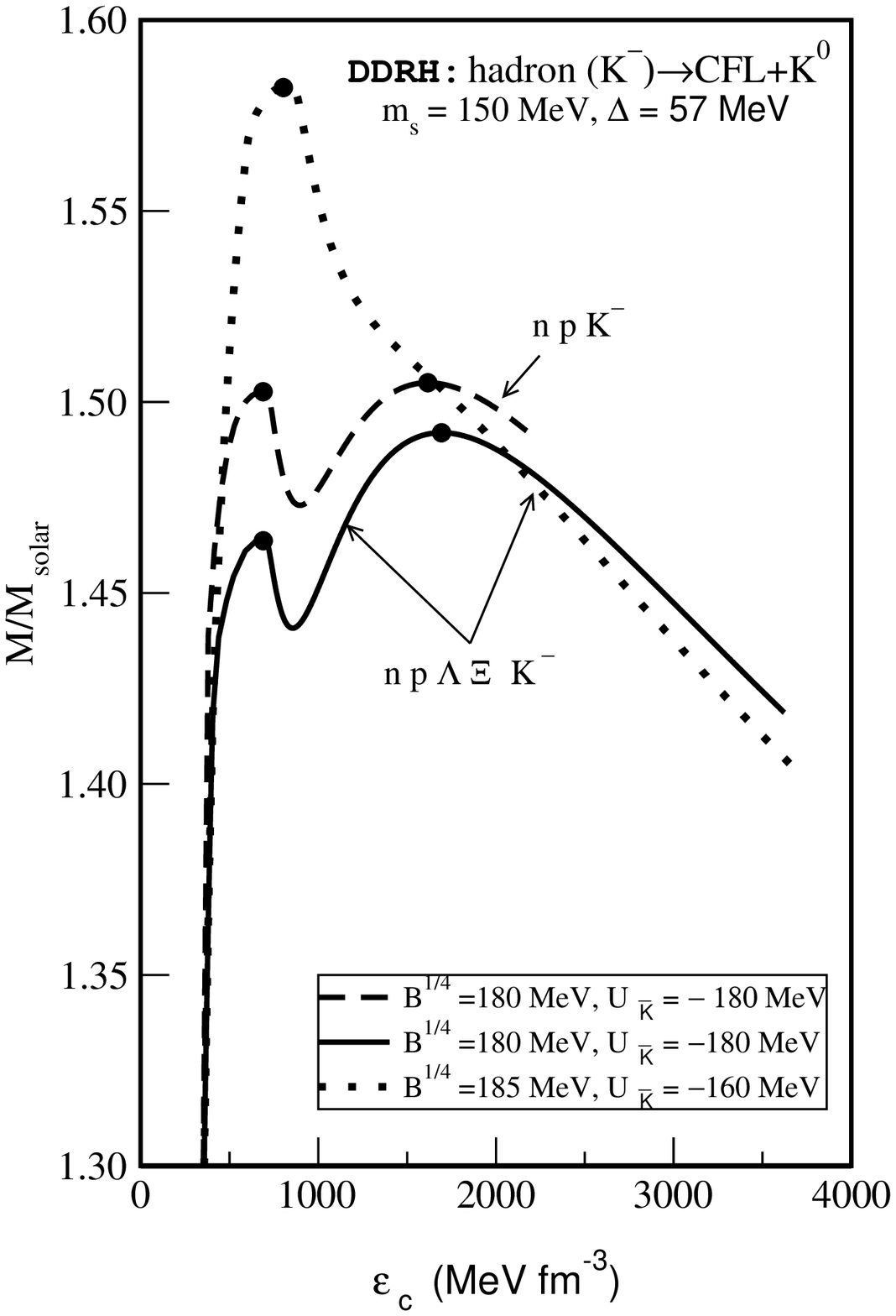,width=16cm,height=19cm}
\caption{The compact star mass sequences are plotted with central energy
density for the
corresponding EoS of Fig. \ref{eosk}.
}\label{mask}
\end{center}
\end{figure}

We also calculate the EoS and structure of compact stars in the phase 
transition from hadronic matter including both hyperons and 
$K^-$ condensate to CFL quark matter. 
Equations of state and compact star mass sequences calculated with
N$\bar K$Q and NH$\bar K$Q matter are plotted in 
Fig. \ref{eosk} and Fig. \ref{mask} 
respectively for various parameter sets. The 
maximum neutron star masses with their central densities are given by
Table II. The dashed line corresponds to N$\bar K$Q matter whereas the other
curves represent NH$\bar K$Q matter in both the figures. 
In Fig. \ref{mask} we find that the static neutron star sequence for 
NH$\bar K$Q matter with $B^{1/4} = 180$ MeV, $\Delta = 57$ MeV, $m_s = 150$ MeV 
and $U_{\bar K}(n_0) = -180$ MeV  
has a lower maximum mass $1.464M_{solar}$ corresponding to central 
energy density $\varepsilon_c=$690.19 ${\em MeV/fm^3}$ than the corresponding 
N$\bar K$Q star of $1.503M_{solar}$ corresponding to central
energy density $\varepsilon_c=$689.59 ${\em MeV/fm^3}$. This may be attributed 
to 
the fact that the EoS in the former case is softer than that of the latter case 
as it is evident from Fig. \ref{eosk}. Here it is interesting to note 
that the maximum mass neutron star including NH$\bar K$Q
matter has a mixed phase core where hyperons, antikaon condensate and CFL
quarks coexist. Similar situation occurs in NH$\bar K$Q star for $\Delta = $ 30
MeV. From Fig. \ref{mask}, we find third 
family solutions beyond neutron star branches in two cases. 
Here superdense stars in the third 
family also contain a pure CFL quark matter core. For the other case studied 
with 
$B^{1/4} = 185$ MeV, $\Delta=57$ MeV, $m_s=150$ MeV and 
$U_{\bar K}(n_0) = -160$ MeV, we
obtain a stiffer EoS resulting in a larger maximum mass star $1.582M_{solar}$ 
corresponding to central energy density $\varepsilon_c=$803.32 
${\em MeV/fm^3}$. 

We have also performed calculation for stars with NH$\bar K$Q matter using a 
larger strange quark mass $m_s = 200$ MeV, $B^{1/4} = 180$ MeV, 
$U_{\bar K} = -180$ MeV and different values of gap. 
For $\Delta = 30$ and 57 MeV, the phase 
transition does not begin even at density as high as 5.2$n_0$
due to the softer EoS resulting from the early appearance of hyperons and 
$K^-$ condensate. And the CFL phase does not occur in the 
corresponding maximum mass neutron star of 1.497 $M_{solar}$ at central density
5.16$n_0$. On the other hand, the CFL phase 
becomes stiffer in the calculation with $\Delta = 100$ MeV and the phase 
transition occurs at low density 1.76$n_0$. Consequently, hyperons and $K^-$
do not appear in the corresponding maximum mass star of 1.639 $M_{solar}$ at
central density 8.81$n_0$. 

\begin{figure} 
\begin{center}
\psfig{figure=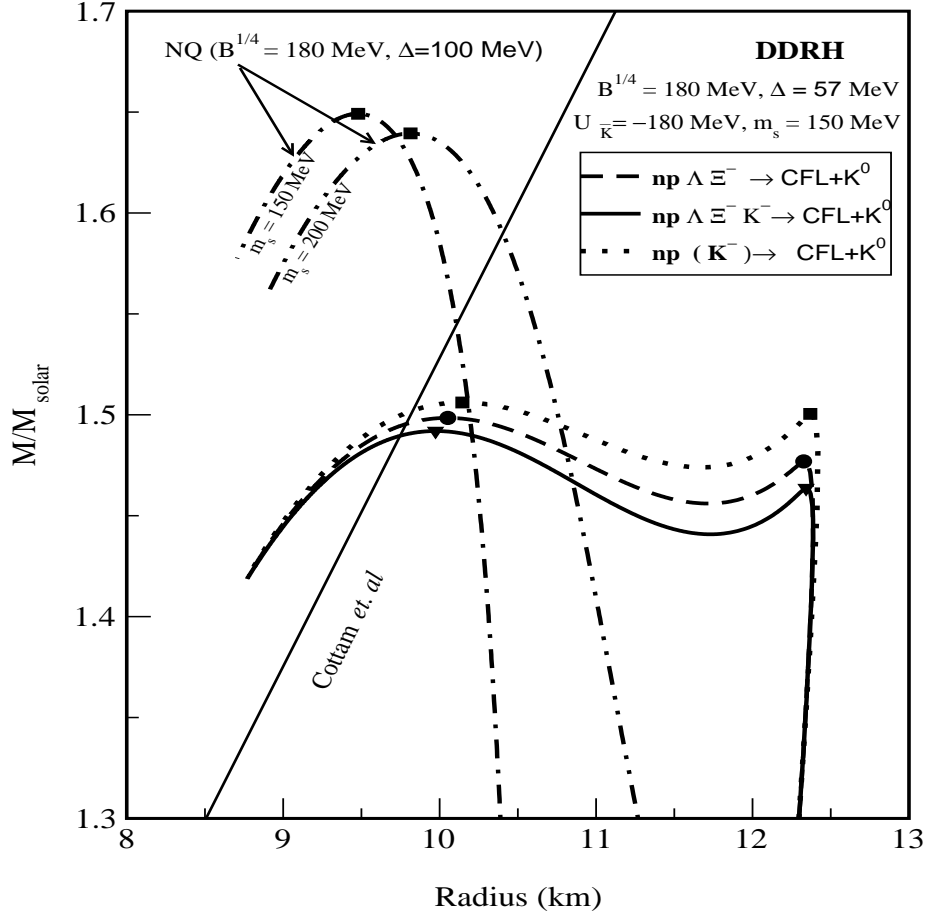,width=16cm,height=19cm}
\caption{The mass-radius relationship for compact star sequences for n, p, 
$\Lambda$, $\Xi$, lepton and CFL+$K^0$  matter with $B^{1/4}= 180$ MeV, 
$m_s$= 150 MeV, $\Delta = 57$ MeV and with $K^-$ ($U_{\bar K}(n_0)=-180$ MeV) 
and 
without $K^-$ condensate. Also, the results for NQ stars with $\Delta = 100$
MeV and different values of $m_s$ are shown here. }\label{masrp}
\end{center}
\end{figure}

The mass-radius relationship of compact stars with NQ, NHQ, N$\bar K$Q and 
NH$\bar K$Q matter for $B^{1/4} = 180$ MeV, $\Delta = 57$ MeV, $m_s = 150$ MeV 
and $U_{\bar K}(n_0) = -180$ MeV are shown in Fig. \ref{masrp}. 
The filled circles, squares and triangles correspond to
maximum masses on the neutron star and third family branch. 
The maximum masses and radii of NQ stars
(dotted line) in the neutron star (third family) branch are 
$1.500(1.506)M_{solar}$ and $12.37(10.15)$ km respectively. 
The presence  of $K^-$ condensate in N$\bar K$Q 
stars does not change maximum masses as it is evident from 
Table \ref{tab2}. The maximum 
masses and their corresponding radii for NH$\bar K$Q 
stars (solid line) in the neutron star 
(third family) branch are 1.464(1.492)$M_{solar}$ and 12.34(9.97) km
whereas those for NHQ stars (dashed line) are 
1.477(1.498)$M_{solar}$ and 12.33(10.05) km respectively. Both
the neutron star and third family branch have smaller
masses in NH$\bar K$Q case than those in NHQ case because 
the EoS are softer in the former case.
It is interesting to note that the radii in the third family branch are 
smaller than their counterparts in the neutron star branch. Also, the 
stars in the third family branch have different compositions than 
those of the neutron star branch. Recently, it has been predicted 
in a perturbative QCD calculation that the maximum mass and radius of the
superdense star in the third family may be $\sim 1 M{\odot}$ and 6 km 
respectively \cite{Fraga}. In Fig. \ref{masrp} we also show the mass-radius 
relationship for NQ stars with  
$B^{1/4} = 180$ MeV, $\Delta = 100$ MeV and $m_s = $ 150 and 200 MeV. The 
maximum masses (radii) of compact stars for $m_s = $ 150 and 200 MeV are
1.649 $M_{\odot}$ (9.48 km) and 1.639 $M_{\odot}$ (9.81 km) respectively. In 
both cases, no third family branch occurs.

From our investigation of compact stars with exotic forms of matter, it is
observed that the thresholds of hyperons, antikaon condensation and CFL quark
phase are sensitive to the hadronic EoS, bag constant, gap and strange quark 
mass. For parameters adopted in our calculation including CFL quark matter, 
maximum neutron star masses
range from 1.464 to 1.649 $M_{solar}$ (Table I and Table II). In the compact 
star with maximum mass 
1.464 $M_{solar}$, all three exotic forms of matter are found to coexist 
whereas the star having a maximum mass 1.649 $M_{solar}$ is 
composed of nuclear+CFL quark matter and has a pure CFL quark matter core.
Similar maximum neutron star masses were found in the calculations including 
only hyperons and antikaon condensate \cite{Bani00,Bani01,Bani02} and also
in the calculation of nuclear-CFL quark matter phase transition \cite{Alf02}.

Now we compare our results with the findings from various observations.
It is argued that those equations of state which predict larger theoretically 
calculated maximum neutron star mass ($M_{max}^{theo}$) than the highest 
measured neutron star mass ($M_{high}^{obs}$), are acceptable \cite{Han03}. 
So far highest accurately measured compact star mass is the Hulse Taylor
pulsar mass which is 1.44 $M_{solar}$ \cite{Tay}. Maximum neutron star masses
obtained in our calculation are found to be larger than the Hulse Taylor 
pulsar mass
i.e. $M_{max}^{theo}>$ $M_{high}^{obs}$ is satisfied by our calculation. 
Recent observations on an isolated neutron star RX J185635-3754 by HST and 
Chandra X-ray observatory and low mass X-ray binary EXO0748-676 by XMM-Newton 
observatory have poured in many interesting data. Walter and Lattimer analysed
HST data from RX J185635-3754 using an atmospheric model \cite{Lat02} and 
predicted radius R = 11.4 $\pm$ 2 km and mass M = 1.7 $\pm$ 0.4 $M_{solar}$. 
This result
is in agreement with the masses and radii of compact stars from our model 
calculation. However, we can not construct any star with radius $\sim 8$ km or 
less as predicted by Drake et al. \cite{Dra} analysing Chandra data on the 
same star. Another possible explanation of Chandra data is given by Zane et al.
\cite{Zan} and it indicates an apparent 
radius $\sim 10-12$ km which is compatible with our soft EoS. 
On the other 
hand, Cottam et al. \cite{Cot} obtained a M-R relationship curve measuring
the gravitational red shift $z = 0.35$ of three
strong spectral lines in X-ray bursts from EXO0748-676. From this observational
M-R relationship, one finds a range of possible values for compact star mass 
1.2-1.8 $M_{solar}$ corresponding to 8-12 km radius. In Fig. \ref{masrp} we 
plot the results of Cottam et al. \cite{Cot}. We note that some of our results 
are consistent with the observational M-R relationship curve. It shows that 
there may be room for exotic forms of matter to coexist in compact stars. As 
the mass of EXO0748-676 is not known, nothing can be concluded with certainty
about dense matter EoS. Further measurements of gravitational redshift and
mass on other neutron stars could put stringent conditions on theoretical 
models for dense matter in compact stars. 

\section{Summary and Conclusions}
We have investigated first order phase transitions from hadronic matter 
including hyperons and antikaon condensate to color-flavor-locked quark matter 
including the condensate of Goldstone boson $K^0$. The 
DDRH model has been adopted here to describe the hadronic phase. This
model takes into account many body correlations in hadronic phase by density 
dependent meson-baryon couplings. Density dependent meson-baryon couplings
are obtained from microscopic Dirac-Brueckner calculations using Groningen
nucleon-nucleon potential. In this calculation meson-(anti)kaon
couplings are density independent. Here $K^-$ condensation is found to be a 
second order phase transition. On the other hand, the CFL quark matter 
is described by the thermodynamic potential of Ref.\cite{Raj01}. 
The role of strange
quark mass and color-superconducting gap on the nuclear-CFL+$K^0$
matter phase transition has been studied for a fixed value of bag constant $B^
{1/4} =180$ MeV. The phase transition is delayed to higher density for
a larger value of $m_s$.
On the other hand, the phase transition sets in earlier as the value of gap 
increases. Also, we have found the early appearance of hyperons and /or
Bose-Einstein condensate of $K^-$ mesons shifts the 
phase transition to higher density. 
 
Equations of state for nuclear matter-CFL+$K^0$ matter phase transition 
have been studied here for different values of $m_s$ and $\Delta$ and for a 
fixed value of bag constant. For a fixed value of gap, 
the EoS becomes stiffer in case of a larger value of $m_s$ resulting in larger 
maximum mass star. For a large value of gap such as $\Delta = 100$ MeV, we find
that different values of strange quark mass have no impact on the EoS and the 
maximum masses of compact stars. 
Similarly, a smaller value of gap softens the EoS leading 
to a smaller maximum mass star. We have also constructed equations of state 
including hyperons and $K^-$ condensate in the hadronic phase for different 
values of bag constant, gap, antikaon optical potential depth and strange 
quark mass. 
The EoS for NHQ, N$\bar K$Q and NH$\bar K$Q matter are softer than that of 
NQ
matter. Consequently, the maximum star masses are smaller in the former cases. 
For all cases studied here, maximum mass neutron stars contain a mixed phase 
core of hadronic+CFL quark matter. There is a window in the parameter space 
for which all three exotic forms of dense matter i.e. hyperons, Bose-Einstein
condensate of $K^-$ mesons and CFL quark matter are found to coexist in 
maximum mass neutron stars. In our calculation with CFL quark matter, maximum 
masses of compact stars
range from 1.464 to 1.649 $M_{solar}$. Our results are consistent with 
the Hulse Taylor pulsar mass and recent observations.  
It is worth mentioning here that we have obtained stable branches of 
superdense 
stars called third family branches beyond neutron star branches for certain 
combinations of parameters. Superdense stars in the third family branch contain
a pure CFL quark matter core. 
The compact stars in the third family have smaller 
radii and different compositions than those of the neutron star branch. 

In this calculation, we do not consider the density dependence of strange 
quark mass, gap and bag parameter. Therefore, it would be interesting to 
investigate how
density dependent strange quark mass, gap and bag parameter in the 
Nambu-Jona-Lasinio model \cite{Stei02} modify our results.

\centerline{\bf Acknowledgements}
D.B. is grateful to Prof. (Dr.) W. Greiner for the warm hospitality of the 
Institut f\"ur Theoretische Physik, Frankfurt Universit\"at where a part of 
the work was completed and the Alexander von Humboldt Foundation for the 
support. 

%
\begin{table}
\caption{Lower($u_l$) and upper($u_u$) boundary of the mixed phase in nuclear-CFL phase transition for different values
of gap $\Delta$ = 0 (unpaired quark matter), 57, 100 MeV
and strange quark mass $m_s$ = 150 and 200 MeV
for a given value of bag constant $B^{1/4} = 180$ MeV. Here 
$u = n_B/n_0$ and saturation density is $n_0=0.18 fm^{-3}$. Average quark 
chemical potential at the onset of phase transition is also given here.
The maximum neutron star masses $M_{max}/M_{solar}$ and their 
corresponding central densities $u_{cent}$=$n_{cent}$/$n_{0}$ are shown below. 
}\label{tab1}
\vskip 0.5 cm
\begin{center}
\begin{tabular}{cccccccc}
$B^{\frac 1 4}$ & $m_s$&$\Delta$&$\mu$ & $u_l$&$u_u$&$u_{cent}$&$\frac {M_{max}}
{M_{solar}}$\\
(MeV)&(MeV)&(MeV)&(MeV)&&\\
\hline  
180 &   150 & 0 &360.03& 1.90& 5.33& 4.77& 1.615\\
180 &  "& 57 & 374.65& 2.14&3.97& 3.97& 1.500\\
180 &  "& 100 &336.07& 1.43&3.09& 9.13& 1.649\\
180 &   200 & 0 &375.06& 2.14& 5.77& 4.51& 1.763\\
180 & " & 57 &480.74& 3.57& 4.61& 4.61& 1.600\\
180 & " & 100 &352.08& 1.76& 3.36& 8.81& 1.639\\
\hline  
\end{tabular}
\end{center}
\end{table}
\newpage
\begin{table}
\caption{The maximum neutron star masses $M_{max}/M_{solar}$ and their 
corresponding central
densities $u_{cent}$=$n_{cent}$/$n_{0}$ with different compositions in a first 
order hadronic to CFL quark matter phase transition. The lower ($u_l$) and 
upper($u_u$) boundary of the mixed phase 
and the threshold density of $K^{-}$ condensation in hadronic phase $u^{K^-}_{th}$, 
where $u = n_B/n_0$, for antikaon optical potential depth
$U_{\bar{K}}(n_0)$ MeV at saturation density $n_{0}=0.18 {\em fm^{-3}}$
are given for various values of bag constant $B^{1/4}$ and gap $\Delta$ and
strange quark mass $m_s = 150$ MeV.
}\label{tab2}
\vskip 0.5 cm
\begin{tabular}{|c|c|c|c|c|c|c|c|c|}
\hline
&$B^{\frac 1 4}$ & $\Delta$& $U_{\bar{K}}(n_0)$&\multicolumn{2}{c|}{CFL}&&
$u_{cent}$&${\frac {M_{max}} {M_{solar}}}$\\
&&&&$u_l$&$u_u$&$u^{K^-}_{th}$&&\\
&(MeV)&(MeV)&(MeV)&&&&&\\
\hline
np$\Lambda\Xi$$\rightarrow$ CFL+$K^0$&180&30&&3.14&4.43&&4.07&1.578\\
  &180&57&&2.27 &3.97&&3.97&1.477\\
  &182&57&&2.60&4.20&&4.06&1.530\\
  &184&57&&3.02&4.48&&4.06&1.569\\
&&&&&&&&\\
np$K^-$$\rightarrow$ CFL+$K^0$&180&30&-180&3.05&4.48&2.18&4.21&1.598\\
          &180&57&-180&2.14&3.97&2.36&3.82&1.503\\
&&&&&&&&\\
np$\Lambda\Xi$$K^-$$\rightarrow$ CFL+$K^0$&180&30&-180&4.73&5.19&2.29&4.87&1.497\\
  &180&57&-180&2.27&3.94&2.68&3.82&1.464\\
  &185&57&-160&3.28&4.63&3.20&4.27&1.582\\
\hline  
\end{tabular}
\end{table}

%

\end{document}